\documentclass[preprintnumbers,prd,showpacs,floatfix,superscriptaddress,nofootinbib,twocolumn, letterpaper]{revtex4-1}
\pdfoutput=1

\usepackage{longtable}
\usepackage{bm}
\usepackage{relsize}
\usepackage{amsfonts}
\usepackage{amsmath}
\usepackage{amssymb,epsf}
\usepackage{latexsym}
\usepackage{graphicx,epsfig}
\usepackage{amssymb}
\usepackage{float}
\usepackage{subfigure}
\usepackage{epstopdf}
\usepackage[colorlinks=true,citecolor=blue,linkcolor=blue,urlcolor=black]{hyperref}
\usepackage{dcolumn}
\usepackage{psfrag}
\usepackage{wrapfig}
\usepackage{makeidx}
\usepackage{epsf}
\usepackage{color}
\usepackage{multirow}
\usepackage{mathtools}

\newcommand{\nn}{\nonumber}
\newcommand{\ph}{\phantom}

\def\InstOfThPhyAstroUG{Institute of Theoretical Physics and Astrophysics,
University of Gdańsk,
80-308 Gda\'{n}sk, Poland}

\def\Tartu{Institute of Physics, University of Tartu, W. Ostwaldi 1, 50411 Tartu, Estonia}

\begin{document}
%%%%%%%%%%%%%%%%%%%%%%%%%%%%%%%%%%%%%%%%%%%%%%%%%%%%%%%%%%%%%%%%%%%%%%%%%%%%

\title{Sudden cosmological singularities in Aether scalar-tensor theories}

\author{João Luís Rosa}
\email{joaoluis92@gmail.com}
\affiliation{\Tartu}
\affiliation{\InstOfThPhyAstroUG}

\author{Tom Zlosnik}
\email{thomas.zlosnik@ug.edu.pl}
\affiliation{\InstOfThPhyAstroUG}

\date{\today}

%%%%%%%%%%%%%%%%%%%%%%%%%%%%%%%%%%%%%%%%%%%%%%%%%%%%%%%%%%%%%%%%%%%%%%%%%%%%
\begin{abstract} 
In this work we analyze the possibility of sudden cosmological singularities, also known as type-II singularities, in the background of a Friedmann-Lemaître-Robertson-Walker (FLRW) geometry in an extension of General Relativity (GR) known as Aether scalar-tensor theories (AeST). Similarly to several scalar-tensor theories, we observe that sudden singularities may occur in certain AeST models at the level of the second-order time derivative of the scale factor. These singularities can either be induced by AeST's scalar field itself in the absence of a fluid matter component, or by a divergence of the pressure component of the fluid. In the latter case, one observes that the second-order time derivative of the scalar field $Q$ is also divergent at the instant the sudden singularity happens. We show that the sudden singularities can be prevented by an appropriate choice of the action and initial conditions, for which a divergence in the scalar field compensates the divergence in the pressure component of the matter fluid, thus preserving the regularity of the scale factor and all its time derivatives. For the models featuring a sudden singularity in the second-order time derivative of the scale factor, an analysis of the cosmographic parameters, namely the Hubble and the deceleration parameters, indicates that cosmological models featuring sudden singularities are allowed by the current cosmological measurements. Furthermore, an analysis of the jerk parameter favours cosmological models that attain a sudden singularity at a faster rate, up to a time of at most $t_s\sim 1.2 t_0$, where $t_0$ is the current age of the universe, and with negative values for the cosmological snap parameter. 
\end{abstract}
%%%%%%%%%%%%%%%%%%%%%%%%%%%%%%%%%%%%%%%%%%%%%%%%%%%%%%%%%%%%%%%%%%%%%%%%%%%%


\maketitle

%%%%%%%%%%%%%%%%%%%%%%%%%%%%%%%%%%%%%%%%%%%%%%%%%%%%%%%%%%%%%%%%%%%%%%%%%%%%
\section{Introduction}\label{sec:intro}
%%%%%%%%%%%%%%%%%%%%%%%%%%%%%%%%%%%%%%%%%%%%%%%%%%%%%%%%%%%%%%%%%%%%%%%%%%%%

The evidence for dark matter is extensive, ranging from the scales of galaxies \cite{Rubin:1980zd} to the largest cosmological scales \cite{Planck:2018vyg}, where it is a vital ingredient of the standard cosmological model. However, evidence for dark matter remains confined to the effect that the gravitational field it sources has on the dynamics of known matter. As such, it remains a possibility that the dark matter effect is due in some part to an interaction between known matter and the gravitational field/spacetime that is not accounted by the theory of General Relativity. 

In 1983 it was discovered by Milgrom \cite{Milgrom:1983ca,Bekenstein:1984tv}  that the then evidence for dark matter in spiral galaxies could alternatively be attributed to either a non-relativistic modification to the gravitational field equation (a non-linear modification to Poisson's equation) or a non-linear relation between the forces acting on a body and its acceleration. The latter possibility (modified inertia) presents a number of challenges to foundational issues such as the construction of an action principle recovering the modified equations of motion \cite{Milgrom:1992hr,Milgrom:2022ifm,Milgrom:2023pmv} and it remains an open question as to whether the idea can be developed to enable predictions to be made on cosmological or regions of strong gravity where post-Newtonian effects become important. 

The former possibility (modified gravity) -  seen as a modification to the Newtonian gravitational field equations - lends itself more readily to embedding into fully-relativistic extensions to General Relativity. A number of such models have been proposed \cite{Bekenstein:1984tv,Bekenstein:2004ne,Sanders:2005vd,Milgrom:2009gv,Babichev:2011kq,Deffayet:2014lba,Woodard:2014wia,Khoury:2014tka,Berezhiani:2015bqa,Blanchet:2015bia,DAmbrosio:2020nev,Avramidi:2023mlc}. In this paper we focus on one particular recent model: the Aether Scalar Tensor (AeST) model \cite{Skordis:2020eui}. This model has the benefit of being able to account well for cosmological data such as the anisotropies in the cosmic microwave background (CMB) radiation and distribution of structure on large scales even in the absence of a dark matter component to the universe. Research into its theoretical and observational consequences is ongoing \cite{Skordis:2021mry,Durakovic:2023out,Verwayen:2023sds,Bataki:2023uuy,Mistele:2023fwd,Mistele:2023paq,Tian:2023gjt,Fu:2023byt,Kashfi:2022dyb}.

Extensions to General Relativity such as AeST generally introduce new degrees of freedom into the gravitational field. Though these degrees of freedom may give rise to an effect similar to that of dark matter in certain regimes, the effect may differ from dark matter in others and so there is scope to distinguish modified gravity and dark matter scenarios experimentally. Modified theories of gravity may also possess pathologies that are not present in dark matter models, thus rendering them potentially non-viable \cite{Seifert:2007fr,Contaldi:2008iw,Babichev:2017lrx,Stahl:2022vaw}.

In the case of the AeST model, the exact form of the model is not fixed by an underlying theoretical framework and so different versions of the model can lead to different phenomenology. An example of this was found in \cite{Rosa:2023qun} where it was shown that a variant of the model certain resembled dark matter (both in the cosmological background and at the level of cosmological perturbations) up to the present day but would invariable lead to a cosmological big rip singularity in the future. 

In this paper we explore the possibility of sudden singularities arising in the AeST model. A sudden singularity happens in cosmological spacetimes whenever at a certain moment in time, time derivatives of order higher than one of metric components diverge. Following Barrow's pioneering work on the subject \cite{Barrow:2004xh,Barrow:2004hk,Barrow:2010wh,Barrow:2010ij,Barrow:2015sga}, sudden singularities were shown to arise also in cosmological models with inhomogeneous equations of state \cite{Dabrowski:2004bz,Trivedi:2022svr} and anisotropic pressures \cite{Barrow:2020rhh}, different types of dark energy models \cite{Nojiri:2005sx,BeltranJimenez:2016dfc,Ghodsi:2011wu}, and in the context of modified theories of gravity \cite{Dabrowski:2009kg,Barrow:2019cuv,Barrow:2020ekb,Rosa:2021ish,Goncalves:2022ggq}. Cosmological tests were performed to assess the viability and impose constraints on models with sudden singularities \cite{Dabrowski:2007ci,Denkiewicz:2012bz,BeltranJimenez:2016fuy}, and their impact on bound systems \cite{Perivolaropoulos:2016nhp}, and it was also shown that quantum effects might delay the sudden singularity \cite{Nojiri:2004ip}. For more information regarding the field of future cosmological singularities, we refer the reader to the following extensive review and references therein \cite{deHaro:2023lbq}.

This manuscript is organized as follows. In Sec. \ref{sec:theory} we introduce the AeST theory of gravity and obtain the corresponding equations of motion and matter distribution in a cosmological framework; in Sec. \ref{sec:sing} we introduce the concept of sudden singularities, analyze the mechanisms via which they can be induced, and prove that it can be prevented in certain particular cases; in Sec. \ref{sec:cosmology} we introduce an explicit cosmological model featuring a sudden singularity and analyze its consequences in terms of the validity of the energy conditions and constraints arising from an analysis of the cosmographic parameters; and in Sec. \ref{sec:concl} we trace our conclusions.

%%%%%%%%%%%%%%%%%%%%%%%%%%%%%%%%%%%%%%%%%%%%%%%%%%%%%%%%%%%%%%%%%%%%%%%%%%%%
\section{Theory and framework}\label{sec:theory}
%%%%%%%%%%%%%%%%%%%%%%%%%%%%%%%%%%%%%%%%%%%%%%%%%%%%%%%%%%%%%%%%%%%%%%%%%%%%

\subsection{Action and field equations}

The AeST theory can be described by an action $S$ of the form
\begin{align}\label{eq:action}
    S=\int \sqrt{-g} \frac{\mathcal L}{2\kappa^2}  d^4x + S_{m}[g],
\end{align}
where $g$ is the determinant of the metric $g_{\mu\nu}$ written in terms of a coordinate system $x^\mu$, $\kappa^2=8\pi \tilde{G}/c^4$ where $\tilde{G}$ is equal to Newton's gravitational constant up to small corrections \cite{Skordis:2020eui} and $c$ is the speed of light, $S_m[g]$ is the matter action, and $\mathcal L$ is the Lagrangian density for the theory, which is given by 
\begin{eqnarray}
      \mathcal L&=&  R-\frac{K_{B}}{2}F_{\mu\nu}F^{\mu\nu}+2(2-K_{B})J^{\mu}\nabla_{\mu}\phi -\\
    &-&(2-K_{B})Y-{\cal F}(Y,Q)- \lambda(g_{\mu\nu}A^{\mu}A^{\nu}+1),\nonumber
\end{eqnarray}
where $A^\mu$, $\phi$ and $\lambda$ are the independent fields of the theory, $K_B$ is a dimensionless constant, $R$ is the Ricci scalar of the metric $g_{\mu\nu}$, and the following quantities were defined:
\begin{eqnarray}
    &&F_{\mu\nu}\equiv 2\partial_{[\mu}A_{\nu]},\\
    &&J^\mu\equiv A^\nu\nabla_\nu A^\mu,\\
    &&Y\equiv\left(g^{\mu\nu}+A^\mu A^\nu\right)\partial_\mu\phi\partial_\nu\phi,\\
    &&Q\equiv A^\mu\partial_\mu \phi,
\end{eqnarray}
where $\partial_\mu$ and $\nabla_\mu$ denote partial and covariant derivatives, respectively. In this work, we use the following convention for index symmetrization and anti-symmetrization:
\begin{eqnarray}
    X_{(\mu\nu)}&\equiv&\frac{1}{2}\left(X_{\mu\nu}+X_{\nu\mu}\right),\nonumber \\
    X_{[\mu\nu]}&\equiv&\frac{1}{2}\left(X_{\mu\nu}-X_{\nu\mu}\right).
\end{eqnarray}
Furthermore, it is useful to define the tensor $q^{\mu}_{\ph{\mu}\nu}= \delta^{\mu}_{\ph{\mu}\nu} + A^{\mu}A_{\nu}$ which projects out the part of a vector field $V^{\mu}$ orthogonal to $A^{\mu}$.
The field equations obtained from varying Eq.\eqref{eq:action} with respect to the fields $g^{\mu\nu},\phi,A_{\mu},$ and $\lambda$ are, respectively:

\begin{widetext}
    \begin{align}
G_{\mu\nu} - K_{B}F_{\mu}^{\ph{\mu}\alpha}F_{\nu\alpha}+(2-K_{B})\big(2J_{(\mu}\nabla_{\nu)}\phi-A_{\mu}A_{\nu}\square\phi 
+2[A_{(\mu}\nabla_{\nu)}A_{\alpha}-A_{(\mu}\nabla_{|\alpha|}A_{\nu)}]\nabla^{\alpha}\phi\big)  \nn\\
- {\cal F}_{Q} A_{(\mu}\nabla_{\nu)}\phi-(2-K_{B}+{\cal F}_{Y})[\nabla_{\mu}\phi \nabla_{\nu}\phi+2Q A_{(\mu}\nabla_{\nu)}\phi] -\lambda A_{\mu}A_{\nu} - \frac{1}{2}g_{\mu\nu}{\cal L} &= 
\kappa^{2} T_{\mu\nu} \label{eq:field}\\
\nabla_{\mu}\left[(2-K_{B})J^{\mu}-(2-K_{B}+{\cal F}_{Y})q^{\mu\nu}\nabla_{\nu}\phi-\frac{1}{2}{\cal F}_{Q}A^{\mu}\right] &=0 \label{eq:eomQ} \\
K_{B}\nabla_{\nu}F^{\nu\mu}+\left(2-K_{B}\right)\left[\left(\nabla^{\mu}A_{\nu}\right)\nabla^{\nu}\phi - \nabla_{\nu}\left(A^{\nu}\nabla^{\mu}\phi\right)\right] -\left[\left(2-K_{B}+{\cal F}_{Y}\right)Q+\frac{1}{2}{\cal F}_{Q}\right]\nabla^{\mu}\phi - \lambda A^{\mu} &=0 \label{eq:vectoreq} \\
g_{\mu\nu}A^{\mu}A^{\nu} +1 &= 0  \label{eq:lambdaeq}
\end{align}
\end{widetext}
where we have defined ${\cal F}_{Y}\equiv\partial{\cal F}/\partial Y$ and ${\cal F}_{Q}\equiv \partial{\cal F}/\partial Q$, and $T_{\mu\nu}$ is the stress energy tensor of matter. 

% Information about the theory up to equations of motion. Please use "eq:field" and "eq:eomQ" as labels to the field equations and equation of motion for Q, respectively, so they are correctly referred to in the sections that follow.

\subsection{Geometry and matter distribution}

In this work we are interested in studying the appearance of sudden singularities in a cosmological context. For this purpose, we assume that the universe is well-described by a homogeneous and isotropic spacetime. These spacetimes are described by the Friedmann-Lemaître-Robertson-Walker (FLRW) line element, which in the usual spherical coordinates $\left(t,r,\theta,\phi\right)$ takes the form
\begin{equation}\label{eq:metric}
ds^2=-dt^2+a^2\left(t\right)\left(\frac{dr^2}{1-kr^2}+r^2d\Omega^2\right),
\end{equation}
where $a\left(t\right)$ is the scale factor of the universe, $k$ is the sectional curvature of the universe which takes the values $\{0,1,-1\}$ for flat, spherical, and hyperbolic geometries, respectively, and $d\Omega=d\theta^2+\sin^2\theta d\phi^2$ is the surface element on the two-sphere. In the following, to preserve the homogeneity and isotropy of the spacetime, all quantities are assumed to depend solely on the time coordinate.

Inserting the metric in Eq. \eqref{eq:metric} into the equation of motion for $\lambda$ given in Eq. \eqref{eq:lambdaeq}, one verifies that the appropriate solution for the vector field $A^\mu$ is
\begin{equation}
    A^{\mu} = \delta^{\mu}_{t} \label{eq:frwa}.
\end{equation}
Following this result, and given that $\phi=\phi\left(t\right)$ to preserve the homogeneity and isotropy of the solution, one obtains 
\begin{align}
Y=0 \\
Q=\dot{\phi}
\end{align}
where a dot $\dot\ $ denotes a derivative with respect to $t$. Furthermore it is useful to define the following function:
\begin{align}
F\left(Q\right) = -\frac{1}{2}{\cal F}(0,Q).
\end{align}

We consider that the matter distribution of the universe is well-described by an isotropic relativistic perfect fluid, for which the stress-energy tensor $T_{\mu\nu}$ takes the diagonal form
\begin{equation}\label{eq:deftab}
T^{\nu}_{\phantom{\nu}\mu}=\text{diag}\left(-\rho,p,p,p\right),
\end{equation}
where $\rho=\rho\left(t\right)$ is the energy density and $p=p\left(t\right)$ is the isotropic pressure of the fluid. The equation for the conservation of energy can be obtained in the usual way by taking a covariant derivative of $T_{\mu\nu}$ as $\nabla_\mu T^{\mu\nu}=0$, where $\nabla_\mu$ denotes the covariant derivative.

Taking Eqs.\eqref{eq:metric} and \eqref{eq:deftab} into the field equations given in Eq.\eqref{eq:field} and the equation of motion for $Q$ given in Eq.\eqref{eq:eomQ}, as well as solving for the field $\lambda$ using the contraction of Eq.\eqref{eq:vectoreq} along $A^{\mu}$\footnote{It may be checked that the components of Eq.\eqref{eq:vectoreq} that are orthogonal to $A^{\mu}$ are trivially satisfied in FLRW symmetry.},  one obtains the following system of cosmological equations:
\begin{equation}\label{eq:cosmo1}
H^2+\frac{k}{a^2}=\frac{8\pi\rho}{3}-\frac{1}{3}\left(F-QF_Q\right),
\end{equation}
\begin{equation}\label{eq:cosmo2}
2\dot H+3H^2+\frac{k}{a^2}=-8\pi p-F,
\end{equation}
\begin{equation}\label{eq:cosmoQ}
F_{QQ}\dot Q+3HF_Q=0,
\end{equation}
where we have defined the Hubble parameter $H=\dot a/a$ and we have adopted a geometrized unit system such that $\tilde{G}=c=1$. On the other hand, the equation for the conservation of energy takes the form
\begin{equation}\label{eq:cosmoT}
\dot\rho+3H\left(\rho+p\right)=0.
\end{equation}
Finally we note that the contributions of the field $Q$ to Eqs. (\ref{eq:cosmo1}) and (\ref{eq:cosmo2}) 
can be rewritten in a more convenient form via the introduction of the following definitions for an energy density $\rho_{Q}$ and pressure $p_{Q}$ as
\begin{align}
\rho_{Q} &= -\frac{1}{8\pi}(F-QF_{Q}),
\label{eq:rhoq}\\
p_{Q} &= \frac{1}{8\pi}F\label{eq:pq}.
\end{align}
Under these definitions, Eq. (\ref{eq:cosmoQ}) takes the form (\ref{eq:cosmoT}) i.e. in FLRW symmetry, the $Q$ field can be recast as a perfect fluid.

It is important to observe that Eqs.\eqref{eq:cosmo1} to \eqref{eq:cosmoT} form a system of four equations out of which only three are linearly independent. This feature can be proved by taking a derivative of Eq.\eqref{eq:cosmo1} with respect to $t$, followed by the use of Eqs.\eqref{eq:cosmoT}, \eqref{eq:cosmoQ}, \eqref{eq:cosmo2} and \eqref{eq:cosmo1} to eliminate the quantities $\dot \rho$, $\dot Q$, $p$ and $\rho$, respectively. This procedure results in an identity, thus proving that the four equations are not linearly independent. One can thus use this fact to discard one of the four equations from the system without loss of generality. Due to its more complicated structure, we chose to discard Eq.\eqref{eq:cosmo2} from the system and work solely with the remaining three equations.

%%%%%%%%%%%%%%%%%%%%%%%%%%%%%%%%%%%%%%%%%%%%%%%%%%%%%%%%%%%%%%%%%%%%%%%%%%%%
\section{Sudden singularities}\label{sec:sing}
%%%%%%%%%%%%%%%%%%%%%%%%%%%%%%%%%%%%%%%%%%%%%%%%%%%%%%%%%%%%%%%%%%%%%%%%%%%%

A sudden singularity, also known as a Type-II future singularity \cite{Nojiri:2005sx} , is defined as an event in the cosmological evolution at which the scale factor $a$, the Hubble parameter $H$ (and consequently the first-order time derivative of the scale factor $\dot a$), and the energy density $\rho$, are all finite but the pressure $p$ is allowed to diverge, thus inducing a divergence in higher-order time derivatives of the scale factor \cite{Barrow:2004xh}. The derivative order at which such a divergence occurs varies among different modified theories of gravity, happening e.g. in second-order time derivatives for Brans-Dicke gravity \cite{Barrow:2019cuv}, third-order time derivatives for $f\left(R,T\right)$ gravity \cite{Goncalves:2022ggq}, or fourth-order time derivatives for hybrid metric-Palatini gravity \cite{Rosa:2021ish}. Let us consider the possibility of sudden singularities arising in in the system of Eqs. \eqref{eq:cosmo1}, \eqref{eq:cosmoQ} and \eqref{eq:cosmoT}. In what follows, we assume that the finite time sudden singularity occurs at some instant $t=t_s$.

\subsection{Sudden singularities induced by the  scalar field}\label{sec:ssvacuum}

Let us start by analyzing the possibility of sudden singularities arising from the contribution of the scalar field $Q$ only, in the absence of the matter fluid component, i.e., we assume $\rho=p=0$. Under this assumption, Eq. \eqref{eq:cosmoT} is identically satisfied, whereas - using the definitions introduced in Eqs. (\ref{eq:rhoq}) and (\ref{eq:pq}), Eqs. \eqref{eq:cosmo1} and \eqref{eq:cosmo2} reduce to 
\begin{equation}\label{eq:cosmo1Q}
H^2+\frac{k}{a^2}=-\frac{1}{3}\left(F-QF_Q\right)\equiv \frac{8\pi}{3}\rho_Q,
\end{equation}
\begin{equation}\label{eq:cosmo2Q}
2\dot H+H^3+\frac{k}{a^2}=-F\equiv -8\pi p_Q,
\end{equation}
\begin{equation}
\label{eq:cosmoTQ}
\dot\rho_Q+3H\left(\rho_Q+p_Q\right)=0.
\end{equation}
Under the assumptions outlined previously, one verifies that in order for Eq.\eqref{eq:cosmo1Q} to be satisfied throughout the entire time evolution, given that $a$ and $H$ remain finite for all times, this implies that $\rho_Q$ must also remain finite, in agreement with the definition of a sudden singularity that keeps the energy density finite. Following the definition of $\rho_Q$ given in Eq. \eqref{eq:cosmo1Q} in terms of the function $F\left(Q\right)$, one verifies that $F-Q F_Q$ must remain finite through the entire time evolution. On the other hand, given that $\dot H$ is allowed to diverge, the validity of Eq. \eqref{eq:cosmo2Q} for the entire time evolution implies that $p_Q$ must also be allowed to diverge, in order to compensate for a possible divergence in $\dot H$. Following the definition of $p_Q$ given in Eq. \eqref{eq:cosmo2Q} in terms of the function $F\left(Q\right)$, one verifies that a divergence in $p_Q$ corresponds to a divergence of the function $F\left(Q\right)$.

The analysis of the previous paragraph implies that the function $F\left(Q\right)$ is potentially divergent at some instant $t_s$, while the combination $F-QF_Q$ must remain finite. For these two conditions to be compatible, the function $F\left(Q\right)$ must behave as $F\sim Q F_Q$ when the time approaches the divergence time $t_s$, i.e., $F\sim Q$, which in turn implies that the scalar field $Q$ diverges at the divergence time. Thus, close to the sudden singularity, the most general form of the function $F\left(Q\right)$ that satisfies these requirements is given by 
\begin{equation}\label{eq:suddenF}
    F\left(Q\right)=c Q+\sum_{n=0}^\infty \frac{a_n}{Q^n}\equiv \sum_{n=-1}^\infty \frac{a_n}{Q^n}, 
\end{equation}
where $c$ and $a_n$ are constant coefficients and $c=a_{-1}$. Replacing the function $F\left(Q\right)$ given in Eq. \eqref{eq:suddenF} into Eq. \eqref{eq:cosmo2Q}, taking the limit $t\to t_s$ and discarding the non-dominant terms, one obtains the following asymptotic relation between the quantities $\ddot a$ and $Q$
\begin{equation}
   - \frac{\ddot a}{a}\simeq\frac{1}{2}c
    Q
\end{equation}
which is valid only near $t=t_s$.

Summarizing, for a model described by any function $F\left(Q\right)$ that admits a series expansion of the form given in Eq. \eqref{eq:suddenF}, sudden singularities induced by a divergence in the scalar field $Q$ could exist at the level of the second order time derivative of the scale factor, i.e., $\ddot a$, while the energy density of the field $\rho_Q$ remains finite.

\begin{figure}
    \centering
    \includegraphics[width=0.45\textwidth]{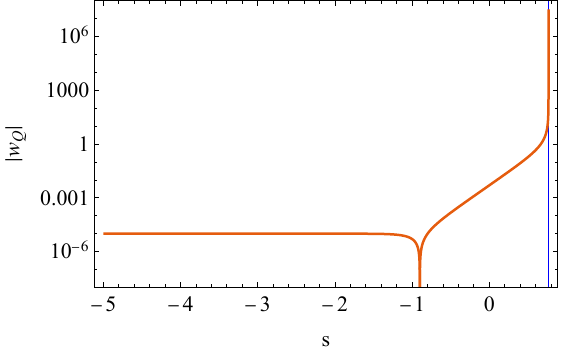}
    \caption{Evolution of the modulus equation of state $w_{Q}=p_{Q}/\rho_{Q}$ for an instance of the model in Eq. (\ref{eq:qfun}) as a function of $s\equiv \mathrm{Log}_{10}a$. For spans of $s$ going from before recombination to the present day, $|w_{Q}|\ll 1$ and it is consistent with bounds on the dark matter equation of state in Generalized Dark Matter models \cite{Ilic:2020onu}. Transition from a small negative value of $w_{Q}$ in the early universe to a diverging positive value at $a=a_{s}$ marks the onset of the sudden singularity (marked by a vertical blue line).}
    \label{fig:wQ}
\end{figure}

\begin{figure}
    \centering
    \includegraphics[width=0.48\textwidth]{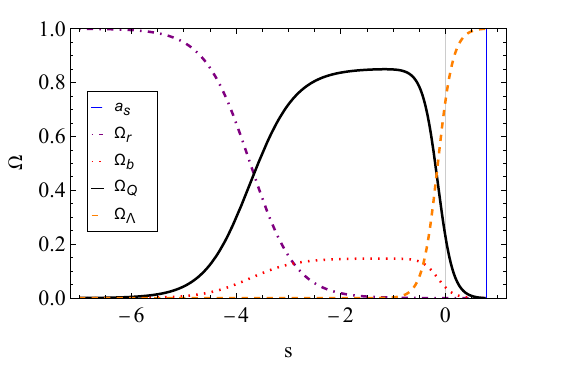}
    \caption{The relative contribution of individual species, namely the radiation, baryonic matter, scalar field, and cosmological constant density parameters $\Omega_r$, $\Omega_b$, $\Omega_Q$, and $\Omega_\Lambda$, respectively, to the overall cosmological density as a function of $s= \mathrm{Log}_{10}(a)$ for the model whose equation of state is shown in Figure \ref{fig:wQ}. The relative abundances satisfy the constraints on the background evolution of the universe \cite{Rosa:2023qun}. At the present moment ($s=0$), the universe has begun entering a period where the expansion is dominated by the cosmological constant. However, unlike the $\Lambda CDM$ model, the illustrated model experiences a sudden singularity in the near cosmic future (the moment of which is denoted by a solid blue line).}
    \label{fig:frwevo}
\end{figure}

Let us now show that allowing for the matter energy density and pressure to be non-vanishing in the universe, there exist forms of $F(Q)$ that resemble dark matter in the cosmological background over a wide span of scale factors and up to the present day, but that evolve to produce a sudden singularity induced by $Q$ in the future. As an example, consider the following function:
\begin{align}
F\left(Q\right) = -e^{-\alpha+\beta Q}- \gamma Q \label{eq:qfun},
\end{align}
where $\alpha$, $\beta$ and $\gamma$ are constant free parameters. The equation of motion in Eq. (\ref{eq:cosmoQ}) can be directly integrated to yield $F_{Q}=-c/a^{3}$, which can be solved for the form of the function introduced in Eq. (\ref{eq:qfun}) to yield:

\begin{align}
Q &= \frac{\alpha -\log \left[\frac{a^3 \beta }{\gamma(a_{s}^{3}-a^3) }\right]}{\beta },
\end{align}
where $a_{s} = (c/\gamma)^{1/3}$.
It follows that

\begin{align}
\rho_{Q} &=  \frac{1}{8\pi }\frac{\gamma\left(a_{s}^{3}-a^3 \right) \left\{\log \left[\frac{a^3 \beta }{\gamma(a_{s}^{3}-a^3) }\right]+1-\alpha \right\}}{a^3 \beta },
\\
p_{Q} &=\frac{1}{8\pi }\bigg\{ \frac{-\left[(\alpha -1) a^3 \gamma \right]+a^3 \gamma  \log \left[\frac{a^3 \beta }{\gamma(a_{s}^{3}-a^3) }\right]-c}{a^3 \beta }\bigg\}.
\end{align}
For positive values of $(c,\gamma,\beta)$, $Q\rightarrow -\infty$ as $a\rightarrow a_{s}$. In this limit $\rho_{Q}\rightarrow 0$ and $p_{Q}\rightarrow +\infty$, signalling a sudden singularity.
For $a\ll (c/\gamma)^{1/3}$, $\rho_{Q}\sim \log(a)a^{-3}$ as shown in Figure \ref{fig:wQ} which illustrates an example of the model in Eq. (\ref{eq:qfun}) where the equation of state for much of cosmic history satisfies $|w_{Q}|\ll 1$ before diverging to $w_{Q}\rightarrow +\infty$ causing a sudden singularity at $a=a_{s}$. Note that the adiabatic sound speed of perturbations $c_{ad}^{2}= \frac{\partial p_{Q}}{\partial \rho_{Q}}$ also diverges to $+\infty$ as $a \rightarrow a_{s}$.
It is noteworthy to mention that the reliability of the function introduced in Eq. \eqref{eq:qfun} extends beyond the vacuum cases and allows one to find cosmological solutions consistent with the current observations. If one follows e.g. a dynamical system analysis under the framework described in Ref. \cite{Rosa:2023qun} in the presence of two relativistic fluids corresponding to baryonic matter and radiation, one obtains the corresponding evolution of the density parameters of the different matter components for this model is given in Figure \ref{fig:frwevo}. One observes that the cosmological evolution presents periods of radiation and matter domination in the past, and is currently undergoing a transition into a cosmological constant dominated epoch. However, unlike it happens for the $\Lambda$CDM model, and even though the cosmological model presented is consistent with the state-of-the-art cosmological measurements by the Planck satellite, this cosmological evolution eventually attains a sudden singularity.

\begin{figure}
    \centering
    \includegraphics[width=0.45\textwidth]{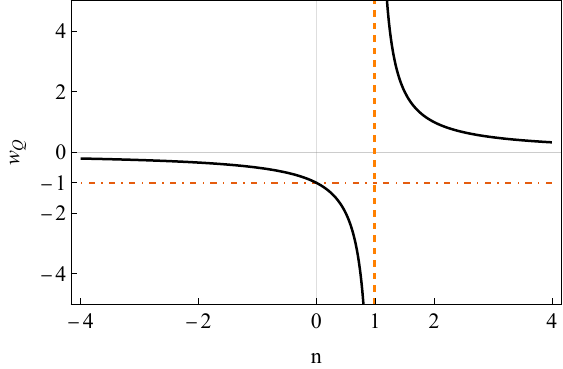}
    \caption{The equation of state of the $Q$ field over for models where $F(Q)\sim Q^{n}$ over some range of $Q$ which is assumed to be dynamically reachable in the sense that it is part of a solution $Q(a)$ which follows from solving Eq.\ref{eq:eomQ}.}
    \label{fig:wQasymptotics}
\end{figure}

Finally, we briefly note that the analysis of Eq. (\ref{eq:suddenF}) can be generalized to the case where, over a dynamically-reached range of $Q$ for which $F(Q) \sim  Q^{n}$, then it follows that its equation of state is $w_{Q}\sim 1/(n-1)$, as illustrated in Figure \ref{fig:wQasymptotics}. As mentioned, the model given in Eq. (\ref{eq:qfun}) can produce a Type II future cosmological singularity, whereas models with $0<n<1$ can produce a Type I/`big rip' future cosmological singularity \cite{Nojiri:2005sx}; an example of this was considered in \cite{Rosa:2023qun} where it was shown that the functional form $F=  k_{0}Q_{0}^{3/2}\sqrt{Q_{0}-Q}$ - for constants $k_{0}$ and $Q_{0}$ - can lead to a Type I singularity despite resembling cold dark matter to a high degree of accuracy up to the present day.

\subsection{Sudden singularities induced by the fluid components}\label{sec:ssfluid}

Consider now an alternative scenario for which the scalar field components are assumed to remain finite, i.e., we assume that the scalar field $Q$ does not contribute to the sudden singularity, while an additional fluid component is present. The aim of this section is to verify if sudden singularities can still arise in the theory even in the presence of a regular scalar field $Q$. The equations that describe this scenario are thus Eqs. \eqref{eq:cosmo1}, \eqref{eq:cosmoQ} and \eqref{eq:cosmoT}.

Following the same procedure as in the previous subsection, one verifies that in order for Eq.\eqref{eq:cosmo1} to be satisfied throughout the entire time evolution, since $a$, $H$ and $\rho$ remain finite, then the quantity $F-QF_Q$ must also remain finite. This regularity can be achieved in two different ways: either the two terms $F$ and $Q F_Q$ remain finite throughout the entire time evolution, or they both diverge at the same rate in such a way that both divergences cancel mutually. The latter option was analyzed in the previous section and it leads to a sudden singularity induced by the scalar field $Q$ for certain forms of the function $F\left(Q\right)$. Thus, in this section, and following the assumption that the scalar field $Q$ is regular, we opt for the first condition, i.e., both the terms $F$ and $QF_Q$ are assumed to remain finite. This assumption also implies that every partial derivative of the function $F$ for any order should remain regular throughout the entire time evolution. We thus infer that the regularity of Eq.\eqref{eq:cosmo1} imposes a regularity of $F$ and, consequently, $Q$.

Having concluded that $F$ and its partial derivatives are regular, along with the scalar field $Q$ itself, one verifies that in order for Eq.\eqref{eq:cosmoQ} to be satisfied throughout the entire time evolution it is necessary that $\dot Q$ remains finite, as otherwise there would be no other divergent term in this equation to preserve its regularity. One thus concludes that $Q$ and its first time derivative $\dot Q$ must be regular throughout the entire time evolution.

Regarding the matter components, if we allow the system to achieve a sudden singularity, the energy density $\rho$ remains finite but the pressure $p$ is allowed to diverge. The only way for Eq.\eqref{eq:cosmoT} to be satisfied throughout the entire time evolution in such a situation is to allow $\dot \rho$ to diverge at the same instant as $p$. In the limit $t\to t_s$, the non divergent terms are subdominant in comparison to the divergent ones, and thus one deduces the asymptotic relation
\begin{equation}\label{eq:suddenT}
\dot\rho \simeq -3Hp,
\end{equation}
an approximation valid only close to $t=t_s$.

Let us now infer how the divergence in $p$ and $\dot\rho$ affects the higher-order time derivatives of the scale factor. This can be done in two ways: either one takes a time-derivative of Eq.\eqref{eq:cosmo1}, or one analyzes directly Eq.\eqref{eq:cosmo2}. These two methods are equivalent, as we have already demonstrated that these two equations are not linearly independent. Since we have previously decided to discard Eq.\eqref{eq:cosmo2} from the system due to this dependence, we shall take the derivative of Eq.\eqref{eq:cosmo1} for this analysis instead. This derivative takes the form
\begin{equation}\label{eq:dcosmo1}
-\frac{\ddot a H}{a}+\frac{k H}{a^2}+H^3=-\frac{4\pi\dot\rho}{3}-\frac{1}{6}Q\dot Q F_{QQ}.
\end{equation}
According to the analysis above, the quantities $a$, $H$, $Q$, $\dot Q$ and $F_{QQ}$ are necessarily finite. Thus, the only way for Eq.\eqref{eq:dcosmo1} to be satisfied throughout the entire time evolution is to allow $\ddot a$ to diverge, in order to counter-balance the divergence in $\dot \rho$. Taking the limit $t\to t_s$ and discarding the sub-dominant terms, one thus obtains the asymptotic relation
\begin{equation}\label{eq:sudden1}
\frac{\ddot a H}{a} \simeq \frac{4\pi\dot\rho}{3},
\end{equation}
which is again only valid near $t=t_s$. We thus conclude that the sudden singularity arises at the second-order time derivative of the scale factor $\ddot a$.

Finally, it is necessary to verify how the sudden singularity affects the higher-order time derivatives of the scalar field $Q$. To do so, we take a time derivative of Eq.\eqref{eq:cosmoQ} from which we obtain
\begin{equation}\label{eq:dcosmoQ}
\left(\frac{\ddot a}{a}-H^2\right)F_Q+H F_{QQ}+\frac{1}{3}\dot Q^2
F_{QQQ}+\frac{1}{3}\ddot Q F_{QQ}=0.
\end{equation}
Similarly to the previous analysis, since $a$, $H$, $Q$, $\dot Q$, and the partial derivatives of $F$ must remain finite, the only way for Eq.\eqref{eq:dcosmoQ} to be satisfied throughout the entire time evolution is to allow for $\ddot Q$ to diverge, to compensate for the divergence in $\ddot a$. Taking the limit $t\to t_s$ and dropping the sub-dominant terms, one obtains the asymptotic relation
\begin{equation}\label{eq:suddenQ}
\frac{\ddot a}{a}F_Q\simeq\frac{1}{3}\ddot QF_{QQ},
\end{equation}
again only valid close to $t=t_s$. The sudden singularity thus also manifest itself in the scalar field $Q$ through its second-order time derivative $\ddot Q$, even though the scalar field $Q$ itself is regular.

The analysis above demonstrates that sudden singularities can arise in this theory even under the assumption that the scalar field $Q$ is regular. These sudden singularities are induced by a divergence in the fluid pressure and they manifest themselves at the level of the second-order time derivatives of the scale factor $\ddot a$ and the scalar field $\ddot Q$. Equations \eqref{eq:suddenT}, \eqref{eq:sudden1} and \eqref{eq:suddenQ} can be rewritten in a more convenient way as to clarify how the divergence in $p$ induces a divergence in $\dot \rho$, $\ddot a$, and $\ddot Q$, depending on the choice of the function $F\left(Q\right)$ that describes the theory, as follows:
\begin{equation}\label{eq:s1}
\frac{\ddot a}{a} \simeq -4\pi p,
\end{equation}
\begin{equation}\label{eq:s2}
\dot \rho \simeq -3Hp,
\end{equation}
\begin{equation}\label{eq:s3}
\ddot Q=\frac{12\pi p F_Q}{F_{QQ}}.
\end{equation}

We note that the analysis conducted in this section, and particularly the behavior obtained in Eq. \eqref{eq:s3}, is valid only in the absence of potentially pathological sets of initial conditions. Indeed, for some particular forms of the function $F(Q)$, it is possible to fine-tune the initial conditions for $Q$ in such a way that Eq. \eqref{eq:s3} ceases to be valid, e.g. cases in which $F_{QQ}$ diverges. These particular cases might induce and/or cancel divergences in $\ddot Q$ that do not directly affect the behavior of the scale factor and are not related to a divergence of the pressure $p$. These divergences thus correspond to mathematical limitations of certain choices of the function $F(Q)$ that are unrelated to the sudden singularity itself and, consequently, are of no interest in the scope of this manuscript. Thus, in order to guarantee that the analysis above is valid independently of the choice of initial conditions, an additional assumption is necessary, i.e., that $F_{QQ}$ remains finite throughout the entire time evolution.

\subsection{Prevention of sudden singularities}

The analysis of the previous two sections indicates that the sudden singularities in AeST theories of gravity can be induced by two different mechanisms: either they are induced by a divergence of the scalar field $Q$, which can happen even in the absence of a fluid component; or they can be induced by the divergence of the pressure component of the fluid, even if the scalar field component is kept regular. It is thus relevant to analyze the hypothesis that these two divergence mechanisms might compensate each other, resulting in a prevention of the sudden singularity and leading to a regular solution for the scale factor up to any order of time derivatives. 

To analyze this hypothesis, we combine the results of the previous two subsections. We assume that both the scalar field and the fluid components are present, which implies that this scenario is described by the set of Eqs. \eqref{eq:cosmo1}, \eqref{eq:cosmoQ}, and \eqref{eq:cosmoT}, the second of which can be rewritten in terms of analogous fluid components as given in Eq. \eqref{eq:cosmoTQ}. One thus obtains the following set of equations, 
\begin{equation}\label{eq:cosmoP1}
H^2+\frac{k}{a^2}=\frac{8\pi}{3}\left(\rho+\rho_Q\right),
\end{equation}
\begin{equation}\label{eq:cosmoPQ}
    \dot\rho_Q+3H\left(\rho_Q+p_Q\right)=0,
\end{equation}
\begin{equation}\label{eq:cosmoPT}
    \dot\rho+3H\left(\rho+p\right)=0.
\end{equation}
Following the procedure of Sec. \ref{sec:ssvacuum}, in order to preserve the regularity of Eq. \eqref{eq:cosmoP1} under the assumption that $H$, $k$, and $\rho$ remain finite throughout the entire time evolution, one concludes that $\rho_Q$ must also remain finite, which again implies that the function $F\left(Q\right)$ must be written in the specific form given in Eq. \eqref{eq:suddenF}. On the other hand, allowing the pressure components $p$ and $p_Q$ to diverge while $H$, $\rho$ and $\rho_Q$ remain finite, Eqs. \eqref{eq:cosmoPQ} and \eqref{eq:cosmoPT} imply that $\dot\rho_Q$ and $\dot\rho$ must diverge at the same instant as $p$ and $p_Q$, respectively. For the purpose of the analysis that follows, we assume that the divergence instant of both $p$ and $p_Q$ is the same, i.e., $t_s$, which can always be done by a choice of appropriate initial conditions. In the limit $t\to t_s$, and discarding the sub-dominant terms, these equations reduce to the two following asymptotic relations
\begin{equation}
    \dot \rho_Q\simeq -3Hp_Q,
\end{equation}
\begin{equation}
    \dot\rho\simeq-3Hp,
\end{equation}
valid close to $t=t_s$. Finally, taking a time derivative of Eq. \eqref{eq:cosmoP1}, taking the limit $t\to t_s$, using the asymptotic relations just obtained for $\dot\rho_Q$ and $\dot \rho$, and discarding the sub-dominant terms, one obtains the asymptotic relation between the second-order derivative of the scale factor and the pressure components as
\begin{equation}\label{eq:prevent}
    \frac{\ddot a}{a}\simeq -4\pi\left(p+p_Q\right).
\end{equation}
This result implies that, since $p$ and $p_Q$ are divergent, in general a sudden singularity at the level of the second-order time derivative of the scale factor is induced. Nevertheless, in the particular case that the pressure components $p$ and $p_Q$ behave asymptotically as $p\simeq -p_Q$ in the limit $t\to t_s$, one observes that the two divergences compensate each other, preventing the sudden singularity. Note that it is not necessary that the two quantities $p$ and $p_Q$ feature the same time-dependent behavior, which would be a strongly constrained scenario, but only that their divergence rates at the singularity time have the same magnitude and opposite signs. Equation \eqref{eq:prevent} can be rewritten in terms of the function $F\left(Q\right)$ through $p_Q$. Inserting the explicit form of $F\left(Q\right)$ given in Eq. \eqref{eq:suddenF} into the result above and requiring that $\ddot a$ is finite, i.e., that the sudden singularity is prevented, one obtains a relationship between $p$ and $Q$ at the singularity time
\begin{equation}\label{eq:preventPQ}
    8\pi p\simeq -c Q.
\end{equation}
We emphasize that it is not necessary that $p$ and $Q$ behave according to the relation above for the entire time evolution, but only close to the divergence time $t\to t_s$. 

The analysis above, where we have considered the first-order time derivative of Eq. \eqref{eq:cosmoP1}, allows one to infer what must be the behavior of the functions $p$ and $p_Q$ such that the sudden singularity is prevented at the level of the second-order time derivative of the scale factor, but such an analysis is not sufficient to guarantee that the sudden singularity is prevented for any order of the time derivatives of the scale factor. To extend this result to higher-order time derivatives of the scale factor, it is necessary to analyze higher-order time derivatives of Eq. \eqref{eq:cosmoP1}, for which such derivatives of the scale factor appear. Following the same procedure as before, one verifies that the higher-order time derivatives of the scale-factor are related to the time derivatives of $p$ and $p_Q$ as
\begin{equation}\label{eq:preventN}
    \frac{a^{(n+2)}}{a}\simeq -4\pi\left(p^{(n)}+p_Q^{(n)}\right),
\end{equation}
where the subscripts $X^{(n)}$ denote the $n$th-order time derivative of the function $X\left(t\right)$. Up to $n=1$, one verifies that in order to prevent a sudden singularity from appearing at the level of the $a^{(n+2)}$ derivative of the scale factor, the behavior of the derivatives $p^{(n)}$ and $p_Q^{(n)}$ for a function $F\left(Q\right)$ given by Eq. \eqref{eq:suddenF} is consistent with Eq. \eqref{eq:preventPQ}, i.e., it can be obtained directly by taking a time derivative of the latter equation. For example, to prevent a sudden singularity in $\dddot a$, it is necessary that $8\pi \dot p \simeq -c\dot Q$  in the limit $t\to t_s$. However, the same is not true for $n>1$, as the derivatives $p_Q^{(n)}$ for $n>1$ depend not only in $Q^{(n)}$ but also in $Q^{(n-i)}$, for every $i\in {1,\cdot,n-1}$. For example, for $n=2$, one has $8\pi \ddot p\simeq -F_{QQ}\dot Q^2-F_Q \ddot Q$. Thus, such a result is only consistent with Eq. \eqref{eq:preventPQ} if the terms proportional to $Q^{(n-i)}$ are subdominant in comparison with the term proportional to $Q^{(n)}$. Given that $Q$ is divergent in the limit $t\to t_s$, and assuming that the field is smooth throughout the entire time evolution, it is an acceptable assumption that every $n$th-order time derivative of the field $Q^{(n)}$ diverges at a larger rate than the derivatives $Q^{(n-i)}$. Thus, the latter terms are subdominant in the limit $t\to t_s$ and consequently Eq. \eqref{eq:preventN} is consistent with Eq. \eqref{eq:preventPQ} at any time derivative order $n$, which implies that the sudden singularity is prevented at every time derivative order of the scale factor. In Appendix \ref{preventionofsuddensingularities} we consider an explicit example of the prevention of a sudden singularity.

The prevention of the sudden singularity analyzed in this section, alongside with the fact that the contributions of the scalar field $Q$ can be conveniently rewritten in the form of a perfect fluid (see Eqs.\eqref{eq:rhoq} and \eqref{eq:pq}), one can merge the matter components into an effective fluid described by $\rho_{\rm eff}=\rho+\rho_Q$ and $p_{\rm eff}=p+p_Q$, and satisfying a conservation equation of the form $\dot\rho_{\rm eff}+3H\left(\rho_{\rm eff}+p_{\rm eff}\right)=0$, in such a way that both the field equations and the conservation equation for the effective fluid remain regular throughout the entire time evolution, thus effectively leading to a system of cosmological equations in which sudden singularities are completely absent.

\section{Sudden singularities in cosmology}\label{sec:cosmology}

In this section, we analyze the physical consequences of having a sudden singularity in a cosmological model, namely we analyze the energy conditions of the model throughout its time evolution and we analyze the constraints the current cosmological observational data imposes on such models. The analysis of this section is mostly model independent, with the exception of the subsection where the energy conditions are analyzed. For the analysis of the energy conditions, we consider the sudden singularities previously analyzed in Sec. \ref{sec:ssfluid}, for which the scalar field $Q$ remains regular and the sudden singularities are induced by the pressure component of the fluid. The remaining analysis of the constraints from the cosmographic parameters is valid for any model for which a sudden singularity appears at the level of the second-order time derivative of the scale factor, independently of the mechanism via which it is induced.

\subsection{Model with a sudden singularity}\label{subsec:model}

The system of Eqs.\eqref{eq:cosmo1}, \eqref{eq:cosmoQ} and \eqref{eq:cosmoT} is an under-determined system of three independent equations for the five unknowns $a$, $\rho$, $p$, $Q$, and $F\left(Q\right)$. This implies that one can impose two extra constraints to determine the system. One of these constraints is naturally an explicit choice of the function $F\left(Q\right)$ as to restrict our analysis solely to adequate models. As for the second constraint, we can directly impose an explicit form for the scale-factor that features a sudden-singularity behavior. Following Barrow \cite{Barrow:2004xh}, let us impose an ansatz for the scale factor of the form

\begin{equation}\label{eq:defa}
a\left(t\right)=\left(a_s-1\right)\left(\frac{t}{t_s}\right)^\gamma+1-\left(1-\frac{t}{t_s}\right)^\delta,
\end{equation}
where $a_s$ is the normalized value of the scale factor at the instant the sudden singularity occurs $t=t_s$, and $\gamma$ and $\delta$ are constant exponents. Note that Eq.\eqref{eq:defa} was chosen in such a way as to guarantee that $a\left(0\right)=0$, i.e., the Big Bang occurs at a time $t=0$. The $n$th-order time derivatives of the scale factor in Eq.\eqref{eq:defa} take the general form
\begin{align}\label{eq:ddefa}
a^{\left(n\right)}\left(t\right)=\left(a_s-1\right)\left(\frac{t}{t_s}\right)^{\gamma-n}t_s^{-n}\prod_{i=0}^{n-1}\left(\gamma-i\right)+\nonumber \\
+\left(-1\right)^{n+1}\left(1-\frac{t}{t_s}\right)^{\delta-n}t_s^{-n}\prod_{i=0}^{n-1}\left(\delta-i\right).
\end{align}
The two terms in Eq.\eqref{eq:ddefa} are responsible for keeping the regularity of $a$ and its derivatives at both the Big Bang $t=0$ and at the sudden singularity $t=t_s$. Indeed, if $\gamma>n$, one verifies that all time derivatives of $a$ up to order $n$ are regular at $t=0$. On the other hand, the sudden singularity appears at $t=t_s$ for the $n$th-order time derivative of $a$ whenever $n>\delta$. Note however that $\gamma$ and $\delta$ should not be whole numbers to avoid the problematic situations $\gamma=n$ and $\delta=n$, for which the derivatives of the scale factor are not well-defined. Since we have proven that in AeST theories the sudden singularity, when present, manifests itself at the level of the second-order time derivative of $a$, we require thus that the exponents $\gamma$ and $\delta$ must satisfy the relation $1<\delta<2<\gamma$.

\begin{figure}
    \centering
    \includegraphics[scale=0.8]{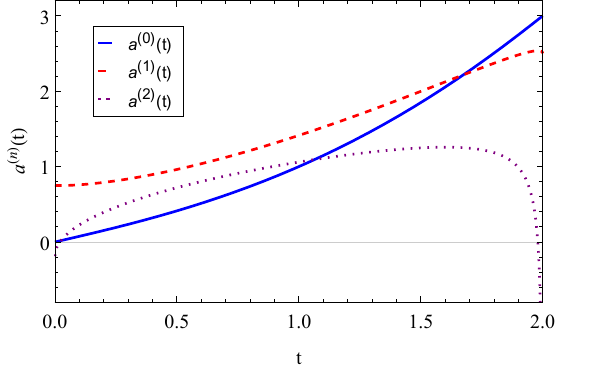}
    \caption{Scale factor $a\left(t\right)$ and its first and second order time derivatives $\dot a$ and $\ddot a$ as a function of time $t$ for the ansatz given in Eq.\eqref{eq:defa} for $a_s=3$, $t_s=2$, $\gamma=2.5$, and $\delta=1.5$.}
    \label{fig:example}
\end{figure}

In Fig. \ref{fig:example} we plot the scale factor $a\left(t\right)$ in Eq.\eqref{eq:defa} as well as its first and second-order time derivatives $\dot a$ and $\ddot a$ for an example model featuring a sudden singularity in $\ddot a$, where we have chosen $a_s=3$, $t_s=2$, $\gamma=2.5$, and $\delta=1.5$. For this model, we verify that all time derivatives of $a$ up to second-order are regular at the origin and throughout the whole time evolution, and that only $\ddot a$ diverges to $\ddot a\to-\infty$ at $t\to t_s$. Also note that $\ddot a$ is negative for small $t$, which implies that there exists a period of decelerated expansion before the accelerated period. Closer to the sudden singularity, $\ddot a$ again changes its sign, inducing a decelerated expansion, just before diverging at the singularity time $t=t_s$. In the particular case of sudden singularities induced by a divergence in the pressure component of the fluid (see Sec. \ref{sec:ssfluid}), from Eqs.\eqref{eq:s1} and \eqref{eq:s2} one verifies that a divergence $\ddot a\to-\infty$ implies that $p$ diverges to $+\infty$ while $\dot \rho$ diverges to $-\infty$. On the other hand, the sign of the divergence in $\ddot Q$ depends on the model chosen for $F\left(Q\right)$, and it follows the sign of the factor $F_Q/F_{QQ}$ at $t\to t_s$. 

\subsection{Energy conditions}

The physical relevance of these models can be addressed via the validity of the energy conditions, a set of conditions the matter fields must satisfy in order to guarantee that certain expected physical properties of matter are verified. For an isotropic perfect fluid, the divergence $p\to+\infty$ while keeping $\rho$ finite guarantees the validity of the Null Energy Condition (NEC) $\rho+p>0$ and the Strong Energy Condition (SEC) $\rho+3p>0$ at the singularity time $t_s$, while the Dominant Energy Condition $\rho>|p|$ is violated. As for the Weak Energy Condition (WEC), described by both the NEC plus the restriction $\rho>0$, must be analyzed separately. 

To obtain the solutions for $\rho$ and $p$ and analyze the energy conditions, it is necessary to select a particular form of the function $F\left(Q\right)$. Integrating the equation of motion for the field $Q$ given in Eq. \eqref{eq:cosmoQ}, one obtains a relation between $F_Q$ and $a$ given by
\begin{equation}
    F_Q=\frac{C}{a^3}, \label{eq:integratedeom}
\end{equation}
for some constant of integration $C$. Upon specifying an explicit form for $F(Q)$, the equation above can be solved directly with respect to $Q$, and afterwards the results can be inserted into Eqs. \eqref{eq:cosmo1} and \eqref{eq:cosmo2} to obtain the solutions for $\rho$ and $p$ as functions of time. These solutions are lengthy, and thus we chose not to write them explicitly. Instead, since we already know that $p\to +\infty$ at the divergence time, it suffices to compute $\rho\left(t_s\right)$ to verify if the WEC is satisfied at the divergence time. 

Let us consider an example of how to apply the reasoning outlined above using a form of the function $F(Q)$ which is known to approximate a family of functions that at later cosmological time provide an alternative to dark matter whilst not in isolation causing a sudden singularity, and analyze the WEC. Consider the following model
\begin{equation}
    F\left(Q\right)=k_0\left(Q-Q_0\right)^2,
\end{equation}
where $k_0$ and $Q_0$ are constant free parameters of the model. Upon making this choice, the solution for the scalar field $Q$ can be obtained directly by integrating Eq. (\ref{eq:integratedeom}) to give

\begin{align}
Q &=  Q_{0} + \frac{C}{2k_{0}a^{3}}.
\end{align}
Following that, the solutions for $\rho$ as a function of $t$ can be obtained by solving Eqs.\eqref{eq:cosmo1}, under the ansatz for the scale factor given in Eq.\eqref{eq:defa}. Taking then the limit $t\to t_s$, the density $\rho$ takes the form

\begin{equation}\label{eq:WEC}
8\pi\rho\left(t_s\right)=\frac{3k}{a_s^2}
-\frac{CQ_{0}}{a_{s}^{3}}-\frac{C^{2}}{4k_{0}a_{s}^{6}}
+\frac{3\gamma^2}{a_s^2t_s^2}\left(a_s-1\right)^2.
\end{equation}

Thus, the WEC is satisfied at the divergence time $t_s$ for any combination of parameters that keeps the right-hand side of Eq.\eqref{eq:WEC} positive. In such a case, the NEC, the WEC and the SEC are all satisfied at the divergence time. 

\subsection{Constraints from the cosmological parameters}\label{sec:sudden}

The observations of the cosmological parameters e.g. from the Planck satellite\cite{Planck:2018vyg}, can be used to impose constraints on the two free parameters of Eq.\eqref{eq:defa}, namely the scale factor $a_s$ at the sudden singularity, and the time $t_s$ at which the divergence occurs. In particular, we are interested in two cosmological parameters, namely the Hubble parameter and the deceleration parameter, which can be written in terms of time-derivatives of the scale factor as
\begin{equation}
H=\frac{\dot a}{a},\qquad q=-\frac{\ddot a a}{\dot a^2}.
\end{equation}
For the scale factor model given in Eq.\eqref{eq:defa}, the cosmological parameters $H$ and $q$ take the forms
\begin{widetext}
\begin{equation}\label{eq:hubble}
H\left(t\right)=\frac{\gamma\left(a_s-1\right)\left(t-t_s\right)\left(\frac{t}{t_s}\right)^\gamma-\delta t\left(1-\frac{t}{t_s}\right)^\delta}{t\left(t-t_s\right)\left[\left(a_s-1\right)\left(\frac{t}{t_s}\right)^\gamma+1-\left(1-\frac{t}{t_s}\right)^\delta\right]},
\end{equation}
\begin{eqnarray}\label{eq:decel}
q\left(t\right)=-\frac{\left(a_s-1\right)\left(\frac{t}{t_s}\right)^\gamma+1-\left(1-\frac{t}{t_s}\right)^\delta}{\left[\gamma\left(a_s-1\right)\left(t-t_s\right)\left(\frac{t}{t_s}\right)^\gamma-\delta t\left(1-\frac{t}{t_s}\right)^\delta\right]^2} \Bigg[\gamma\left(\gamma-1\right)\left(a_s-1\right)\left(t-t_s\right)^2\left(\frac{t}{t_s}\right)^\gamma-\delta\left(\delta-1\right)t^2\left(1-\frac{t}{t_s}\right)^\delta\Bigg].
\end{eqnarray}
\end{widetext}

These parameters $H$ and $q$ have been measured experimentally and their current observed values at the present time are $H\left(t_0\right)\equiv H_0\sim67.66 \text{km s}^{-1}\text{Mpc}^{-1}\sim 2.19\times 10^{-18}\text{s}^{-1}$ and $q\left(t_0\right)\equiv q_0\sim-0.53$, where the present time $t_0$, also referred to as the age of the universe, is $t_0\sim 13.79 \text{Gy}\sim 4.35\times 10^{17}\text{s}$. Inserting the measured values of the cosmological parameters $H_0$, $q_0$ and $t_0$ into Eqs.\eqref{eq:hubble} and \eqref{eq:decel} allows one to write a system of two equations $H\left(t=t_0\right)=H_0$ and $q\left(t=t_0\right)=q_0$ for the four unknowns $a_s$, $t_s$, $\delta$ and $\gamma$. Since the exponents $\gamma$ and $\delta$ must satisfy the inequalities $1<\delta<2<\gamma$ to guarantee the appearance of the sudden singularity at the right order of the time derivatives of the scale factor, see Sec. \ref{subsec:model}, a possible way to solve this system is to choose the values of $\gamma$ and $\delta$ \textit{a priori} and solve the two equations for $a_s$ and $t_s$, if any solutions exist for that combination of $\gamma$ and $\delta$. This method allows one to predict the divergence time $t_s$ for these models, as well as the size of the scale factor $a_s$ at the divergence time, while keeping the cosmological parameters consistent with their observational values.

\begin{figure}
    \centering
    \includegraphics[scale=0.8]{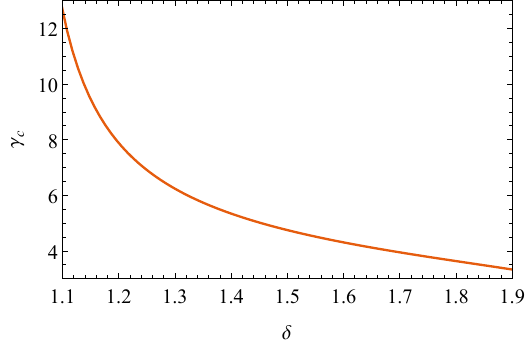}
    \caption{Critical value $\gamma_c$ as a function of the exponent $\delta$.}
    \label{fig:gamma}
\end{figure}

\begin{figure}
    \centering
    \includegraphics[scale=0.8]{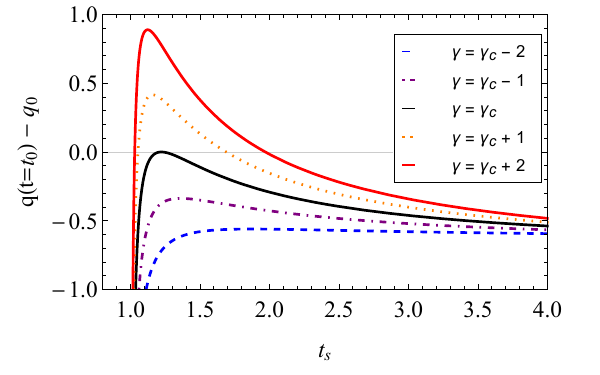}
    \caption{Combination $q\left(t_0\right)-q_0$ as a function of the divergence time $t_s$ for $\delta=1.5$ and for different values of $\gamma$. The critical value of $\gamma$ for this combination of parameters is $\gamma_c\sim 4.74789$}
    \label{fig:critical}
\end{figure}

Unlike in the cases of models with sudden singularities arising at the third or fourth order time derivatives of the scale factor \cite{Rosa:2021ish,Goncalves:2022ggq}, not all possible combinations of $\gamma$ and $\delta$ for which the ansatz in Eq.\eqref{eq:defa} develop a sudden singularity at the second-order time derivative of the scale factor allow for choices of $a_s$ and $t_s$ that produce models consistent with the cosmological observations. Solving Eq.\eqref{eq:hubble} with respect to $a_s$, replacing the result into Eq.\eqref{eq:decel}, and setting a value for $\delta$, the resultant equation can feature zero, one, or two possible solutions for $t_s$ depending on the value of $\gamma$. Indeed, one verifies that there exists a critical value of $\gamma$, say $\gamma_c\left(\delta\right)$, such that if $\gamma<\gamma_c$ no value of $t_s$ is consistent with the cosmological parameters, if $\gamma=\gamma_c$ a single value of $t_s$ is consistent with the cosmological parameters, and if $\gamma>\gamma_c$ there are up to two values of $t_s$ that produce models consistent with the cosmological parameters. The value of $\gamma_c$ as a function of $\delta$ is given in Fig. \ref{fig:gamma}, where one can observe that $\gamma_c$ decreases monotonically with an increase in $\delta$. To clarify this situation, in Fig. \ref{fig:critical} we plot $q\left(t=t_0\right)-q_0$ as a function of $t_s$ for $\delta=1.5$ and different values of $\gamma$, where $\gamma_c\sim 4.74789$. The zeroes of the function $q\left(t=t_0\right)-q_0$ correspond to the values of $t_s$ that produce models consistent with the observed cosmological parameters. 


\begin{figure*}
    \centering
    \includegraphics[scale=0.66]{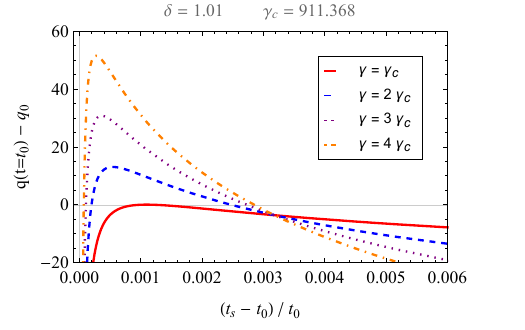}
    \includegraphics[scale=0.66]{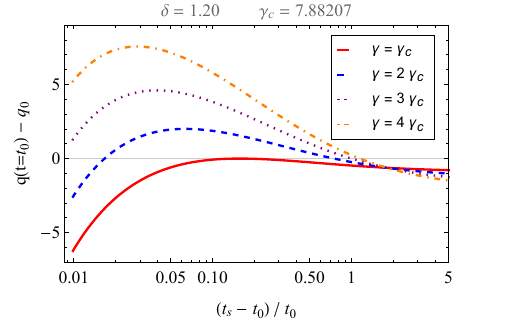}
    \includegraphics[scale=0.66]{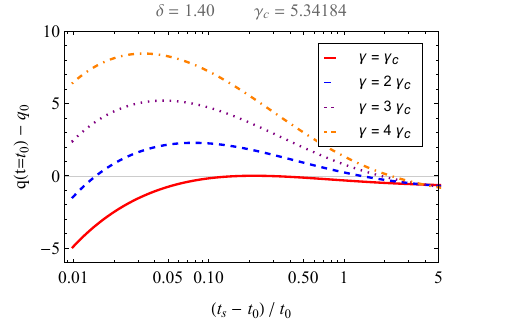}\\ 
    \ \\
    \includegraphics[scale=0.66]{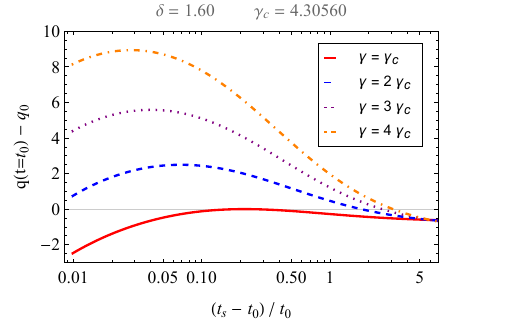}
    \includegraphics[scale=0.66]{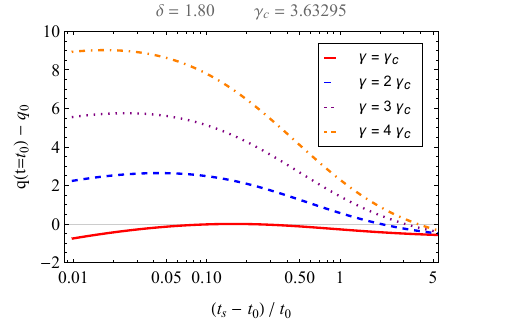}
    \includegraphics[scale=0.66]{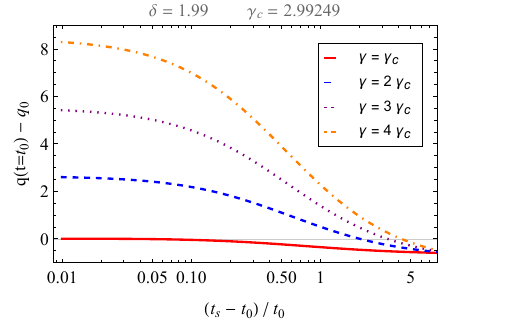}
    \caption{Combination $q\left(t_0\right)-q_0$ as a function of the normalized divergence time $\left(t_s-t_0\right)/t_0$ for different values of $\delta$ and $\gamma$.}
    \label{fig:times}
\end{figure*}

To illustrate the behavior of the value of $t_s$ as a function of $\gamma$ and $\delta$ in the parameter space for which solutions consistent with the cosmological observations exist, we have chosen a set of six different values of $\delta$, namely $\delta=\{1.01; 1.2; 1.4; 1.6; 1.8; 1.99\}$. For Each of these values, we have computed the critical value $\gamma_c$ and plotted the quantity $q\left(t=t_0\right)-q_0$ as a function of $\left(t_s-t_0\right)/t_0$ for four different values of $\gamma$, namely $\gamma=n \gamma_c$, with $n\in\{1,2,3,4\}$. The results are depicted in Fig. \ref{fig:times}. One verifies that for $\gamma \gtrsim \gamma_c$ the system presents two possible solutions for $t_s$ consistent with the cosmological observations, say $t_s^{(1)}$ and $t_s^{(2)}$ with $t_s^{(1)}<t_s^{(2)}$, these two solutions degenerating into a single one at $\gamma=\gamma_c$, i.e., $t_s^{(1)}=t_s^{(2)}$. As one increases $\gamma$, $t_s^{(1)}$ decreases and $t_s^{(2)}$ increases, and eventually $t_s^{(1)}=t_0$ for a large enough $\gamma$. If one further increases $\gamma$, a single solution for $t_s$ consistent with the observed cosmological parameters exists, and it is given by $t_s^{(2)}$. 

\begin{figure*}
    \centering
    \includegraphics[scale=0.8]{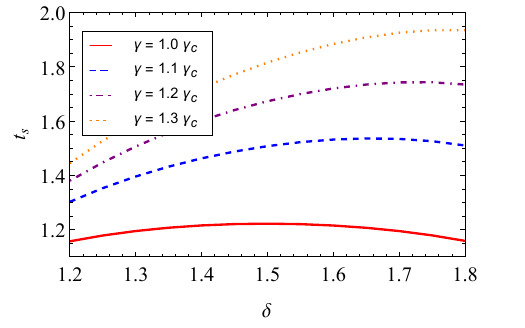}
    \includegraphics[scale=0.8]{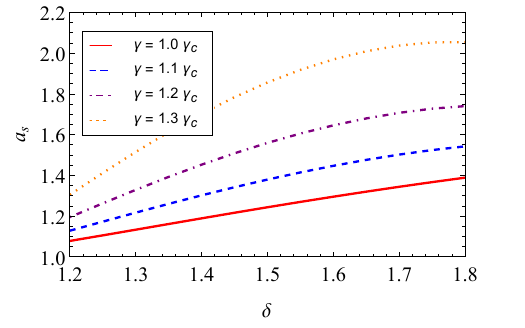}
    \caption{Normalized divergence time $t_s^{(2)}$ (left panel) and the corresponding normalized divergence scale factor $a_s$ (right panel) as functions of $\delta$ for different values of $\gamma\gtrsim \gamma_c$.}
    \label{fig:sudden}
\end{figure*}

Although the solutions $t_s^{(1)}$ and $t_s^{(2)}$ produce cosmological models consistent with the current observations for $t_0$, $H_0$, and $q_0$, the two models obtained can be distinguished by recurring to higher-order cosmological parameters, namely the cosmological jerk $j$ and snap $s$ parameters, defined in terms of the higher-order derivatives of the scale factor as
\begin{equation}\label{eq:defjs}
j=\frac{\dddot a a^2}{\dot a^3}, \qquad s=\frac{a^{(4)} a^3}{\dot a^4}.
\end{equation}
The cosmological parameters $j$ and $s$ have not been measured experimentally, although some studies indicate that the current observational status favours a small value of the jerk parameter, or order $\sim 1$ \cite{Mukherjee:2020ytg,FaisalurRahman:2021blt,AlMamon:2018uby,Mehrabi:2021cob,Demianski:2016dsa,Zhai:2013fxa}. For the combinations of $\delta$ and $\gamma$ for which the two solutions $t_s^{(1)}$ and $t_s^{(2)}$ exist, we have observed that the measurements for the present cosmological jerk parameter $j\left(t=t_0\right)\equiv j_0$ quickly achieve values several orders of magnitude above unity, in conflict with the studies mentioned above. Thus, we chose to discard the solutions $t_s^{(1)}$ from the analysis and consider solely the solution $t_s^{(2)}$. In Fig. \ref{fig:sudden}, we plot the normalized divergence time $t_s^{(2)}$ and the corresponding normalized divergence scale factor $a_s$ as functions of $\gamma$ for different values of $\gamma\gtrsim \gamma_c$. We observe that an increase in $\gamma$ results in an increase in both $t_s$ and $a_s$, and also that $a_s$ increases monotonically with $\delta$ in the range considered, while $t_s$ presents a global maximum at some value of $\delta$, for a constant $\gamma$. 

\begin{figure*}
    \centering
    \includegraphics[scale=0.8]{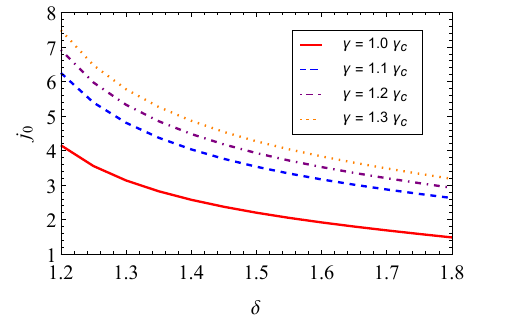}
    \includegraphics[scale=0.8]{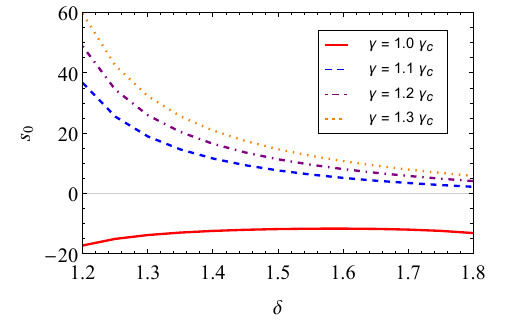}
    \caption{Present jerk parameter $j_0$ (left panel) and snap parameter $s_0$ (right panel) as functions of $\delta$ for different values of $\gamma\gtrsim \gamma_c$.}
    \label{fig:jerksnap}
\end{figure*}

Let us now consider the present cosmological parameters $j_0$ and $s_0$. In Fig. \ref{fig:jerksnap} we plot the present values of the jerk parameter $j_0$ and the snap parameter $s\left(t=t_0\right)=s_0$ obtained by considering the solutions for $t_s$ and $a_s$ consistent with the cosmological observations, and for the same range of the parameters $\delta$ and $\gamma$. One observes that $j_0$ is always positive and it increases with an increase in $\gamma$ and a decrease in $\delta$. On the other hand, the snap parameter $s_0$ remains negative for $\gamma\sim\gamma_c$ independently of the value of $\delta$, but it turns positive for larger values of $\gamma$. When positive, the snap parameter decreases with an increase in $\delta$. An interesting result of this analysis is that the jerk parameter $j_0$ quickly scales to large values compared to unity with small variations of $\gamma$, and for values of delta close to the lower bound $\delta\sim 1$. This implies that cosmological models with a value of $\delta$ close to the upper bound $\delta\sim 2$ and a value of $\gamma$ close to $\gamma_c$ are favored by this analysis, which in turn constrains the expected values of the snap parameter $s_0$ to be negative and of the order of $\sim -10$. In the future, once the values of $j_0$ and $s_0$ have been measured, those measurements can be used to impose constrains on the values of $\gamma$ and $\delta$, potentially leading to a narrowing of the models consistent with the cosmological observations, or even completely excluding the possibility of a future sudden singularity under the framework considered.


We note that the analysis conducted in this section, although motivated by the fact that the AeST theories can feature sudden singularities at the level of the second-order time derivative of the scale factor $\ddot a$, is completely theory independent, as no assumptions concerning the explicit form of the theory were considered. This means that the analysis holds for any theory of gravity that also presents sudden singularities at the level of $\ddot a$, as long as Eq.\eqref{eq:defa} is a solution of the field equations, which includes e.g. certain particular forms of scalar-tensor and $f\left(R\right)$ theories of gravity.

%%%%%%%%%%%%%%%%%%%%%%%%%%%%%%%%%%%%%%%%%%%%%%%%%%%%%%%%%%%%%%%%%%%%%%%%%%%%
\section{Conclusions}\label{sec:concl}
%%%%%%%%%%%%%%%%%%%%%%%%%%%%%%%%%%%%%%%%%%%%%%%%%%%%%%%%%%%%%%%%%%%%%%%%%%%%

In this work we have explored the possibility of cosmological models featuring sudden future singularities arising in AeST theories of gravity. These singularities are characterized by a divergence of the time derivatives of the scale factor at a certain order higher than one, while the scale factor itself and its first-order time derivatives remain finite. Similarly to what happens with $f\left(R\right)$ gravity and some forms of scalar-tensor theories of gravity e.g. Brans-Dicke gravity, we have shown that sudden singularities in AeST theories may arise at the level of the second-order time derivative of the scale factor, and that they are induced either by a divergence of derivatives of the scalar field of the model, or by a divergence in the pressure component of a relativistic fluid.

The analysis of the modified field equations, along with the equation of motion for the scalar field $Q$, close to the instant at which the sudden singularity arises shows that the pressure divergence induces also a divergence in the first-order time derivative of the energy density and the second-order time derivative of the scalar field $Q$. Furthermore, we have shown that the divergence in $\ddot a$, as well as the divergence in $\dot \rho$, occur at negative values, i.e., $\ddot a\to -\infty$ and $\dot\rho \to -\infty$, while $p\to +\infty$. The positivity of $p$ at the divergence time guarantees that the NEC and the SEC are always satisfied at the divergence time $t_s$, whereas the DEC is always violated. The WEC may or may not be satisfied depending on the form of the function $F\left(Q\right)$. Nevertheless, it seems to be always possible to fine-tune the parameters of the theory in such a way that the WEC is satisfied. 

Even in the absence of a fluid component, we have shown that the scalar field $Q$ itself may induce a sudden singularity at the level of the second-order time derivative of the scale factor. Furthermore, we have shown that there exist specific functional forms for $F(Q)$ which provide a viable alternative to dark matter at the level of the cosmological background and can produce a realistic expansion history in the presence of other matter despite leading to a sudden singularity in the cosmological future.

An important consequence of the ability of $Q$ to produce sudden singularities in the absence of matter is that, in the presence of a fluid component, if the pressure associated to the scalar field diverges at the same rate as the pressure component of the fluid but with an opposite sign, these two divergences can compensate each other, thus preserving the regularity of $\ddot a$ and preventing the sudden singularity. This relationship between the two pressure components is only necessary as one approaches the divergence time, while no restrictions on the behavior of these functions are necessary far from that instant. Furthermore, such a cancellation of divergences may only occur if the function $F\left(Q\right)$ behaves asymptotically as $F\sim F_Q Q$, close to the divergence time.

Using the current measurements of the cosmological parameters, namely the Hubble parameter $H_0$, the deceleration parameter $q_0$ and the age of the universe $t_0$, we constrained the parameters of the cosmological models featuring sudden singularities and obtained predictions for the instant $t_s$ at which the sudden singularity occurs and the corresponding value of the normalized scale factor $a_s$. Furthermore, this analysis allowed for the prediction of the current values of higher-order cosmographic parameters, namely the jerk $j_0$ and the snap $s_0$ parameters. In order to keep $j_0\sim 1$, consistently with several other studies (see Sec. \ref{sec:sudden}), one expects the cosmological snap parameter to be of order $\sim -10$, while the divergence time and divergence scale factors are expected to be of the order $t_s\sim 1.2 t_0$ and $a_s\sim 1.4 a_0$. We note that this analysis is theory independent, and thus it is also applicable to any other theory of gravity for which sudden singularities may arise at the order of $\ddot a$, e.g. the ones mentioned above.

In the future, with the gathering of more precise cosmographic data, we expect the values of the present jerk and snap parameters to be more accurately measured. Such measurements would allow one to further constrain the parameter space of the cosmological models featuring sudden singularities, potentially leading to a more precise prediction of the divergence time and divergence scale factor, or eventually leading to the complete exclusion of this scenario under the framework considered.

Finally we comment on what these results mean for the viability of the AeST model as an alternative to dark matter. 
The theory inevitably possesses non-canonical kinetic terms for the scalar field $\phi$ but, at the moment, there are no foundational principles for the exact form that these terms take. It should be noted that non-canonical kinetic terms for scalar fields can arise in fundamental physics \cite{Sen:2002an,Lambert:2002hk}. The utility of exploring cosmological solutions is that they can provide an efficient mechanism for uncovering which kinetic terms can lead to pathological behaviour. It has now been shown that models exist that can be entirely consistent with cosmological data up to the present moment but that would lead to a sudden singularity or big rip singularity in the cosmic future. That such events must take place in our cosmological future places constraints on the theory. It remains an open question as to whether their potential existence at any cosmic time signifies a deeper problem.

%%%%%%%%%%%%%%%%%%%%%%%%%%%%%%%%%%%%%%%%%%%%%%%%%%%%%%%%%%%%%%%%%%%%%%%%%%%%
\begin{acknowledgments}
J.L.R. acknowledges the European Regional Development Fund and the programme Mobilitas Pluss for financial support through Project No.~MOBJD647, Fundação para a Ciência e Tecnologia through project number PTDC/FIS-AST/7002/2020, and Ministerio de Ciencia, Innovación y Universidades (Spain), through grant No. PID2022-138607NB-I00. This work is part of the project No.~2021/43/P/ST2/02141 co-funded by the Polish National Science Centre and the European Union Framework Programme for Research and Innovation Horizon 2020 under the Marie Sklodowska-Curie grant agreement No. 945339.
\end{acknowledgments}

\appendix
\section{Prevention of sudden singularities}
\label{preventionofsuddensingularities}
\begin{figure}[h]
    \centering
    \includegraphics[width=0.45\textwidth]{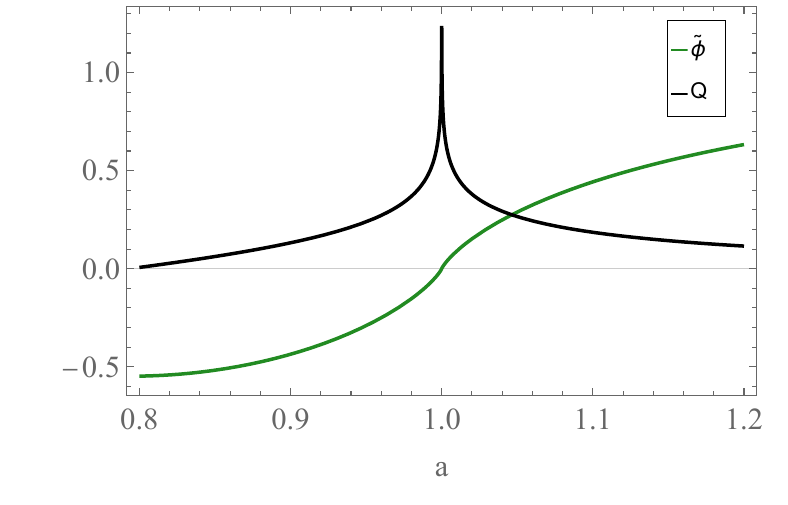}
    \caption{The evolution of $Q$ and $\tilde{\phi}$ as a function of $a$ for a case of the model (\ref{oddf}).}
    \label{fig:HiggenSig}
\end{figure}

In this section we provide an explicit example of a function $F(Q)$ which, in isolation, would cause a sudden singularity but, in the presence of another fluid which cancels out the diverging terms in the stress energy tensor, does not result in a sudden singularity. Furthermore, it can be evolved in time past the point where divergences in the pressure of the individual components occurs. Consider the following function:
\begin{align}
F(Q) &= -2\sqrt{ e^{-\alpha+\beta Q}}- \gamma Q. \label{oddf}
\end{align}
From the integrated scalar field equation of motion we then have that
\begin{align}
\frac{C}{a^{3}}+\gamma + \frac{1}{2}\beta \sqrt{e^{-\alpha + \beta Q}} &=0.
\end{align}
Taking example values $(C=1,\gamma=\beta=-1,\alpha=0)$ we obtain have the solution
\begin{align}
Q &= -\ln\bigg[\bigg(\frac{1}{a^{3}}-1\bigg)^{2}\bigg].
\end{align}
If we take $a=1$ to occur at a moment when the universe is dominated by the cosmological constant (as is expected to happen in our own universe in the future), then we have $H \approx \sqrt{\Lambda/3}$ and so $d\phi/da = d\phi/dt / (da/dt) \approx \sqrt{\frac{3}{\Lambda}}Q/a$. Defining a rescaled scalar field $\tilde{\phi}=\phi\sqrt{\Lambda/3}$, we plot $\tilde{\phi}(a)$ in Figure  \ref{fig:HiggenSig}.

The results above indicate that there is a solution corresponding to a continuous $\tilde{\phi}(a)$. Care should be taken in its interpretation - it is obtained by taking the positive square root of $e^{-Q}$ for $a \leq 1$ and the negative square root of $e^{-Q}$ for $a \geq  1$, and matching the solutions at $a=1$. 

Recall that the pressure associated with the degree of freedom $F$ is $p_{Q} = F/8\pi$ and so this contribution to the stress energy tensor diverges as $a\rightarrow 1$. If another degree of freedom $f$ is present that acts as a perfect fluid in the manner that the model based on the field $Q$ does, then it can be tuned to evolve so as to have a pressure $p_{f}$ that cancels the diverging term from $p_{Q}$ but that can nonetheless be evolved `through' the singular point $a=1$ via its equations of motion. This scenario is undoubtedly highly fine tuned and it is not clear whether the model (\ref{oddf}) possesses pathologies when considered in other situations; however the example serves as an illustration that degrees of freedom that in isolation would produce a sudden singularity can, in the presence of other degrees of freedom, lead to avoidance of the sudden singularity and evolution of the universe to later times.

%%%%%%%%%%%%%%%%%%%%%%%%%%%%%%%%%%%%%%%%%%%%%%%%%%%%%%%%%%%%%%%%%%%%%%%%%%%%

%%%%%%%%%%%%%%%%%%%%%%%%%%%%%%%%%%%%%%%%%%%%%%%%%%%%%%%%%%%%%%%%%%%%%%%%%%%%

\end{document}